\documentclass[aps,pre,twocolumn,float]{revtex4-1}
\usepackage[normalem]{ulem}
\usepackage{amsmath}
\usepackage{amssymb}
\usepackage{color}
\usepackage{graphicx}
\usepackage{natbib}
\usepackage{CJK}

\usepackage{xcolor,hyperref}
\hypersetup{
   colorlinks,
   linkcolor={blue!50!black},
   citecolor={blue!50!black},
   urlcolor={blue!80!black}
}

\DeclareMathAlphabet{\mathitbf}{OML}{cmm}{b}{it}

\definecolor{pink}{rgb}{1,1,0} 
\definecolor{red}{rgb}{1,0,0}  
\definecolor{blue}{rgb}{0,0,1} 
\definecolor{green}{rgb}{0,1,0}
\definecolor{yellow}{rgb}{1,1,0}
\definecolor{orange}{rgb}{1,0.5,0}
\definecolor{white}{rgb}{1,1,1}

\begin{document}

\title{Critical yielding rheology: from externally deformed glasses to active systems}

\author{Carlos Villarroel${}^{1}$, Gustavo D\"uring${}^{1,}$${}^{2}$  }
\affiliation{${}^1$Instituto de F\'isica, Pontificia Universidad Cat\'olica de Chile, Casilla 306, Santiago, Chile\\ ${}^2$ANID - Millenium Nucleus of Soft Smart Mechanical Metamaterials, Santiago, Chile}


\begin{abstract} We use extensive computer simulations to study the yielding transition under two different loading schemes: standard simple shear dynamics, and self-propelled, dense active systems.  In the active systems a yielding transition toward a liquid phase is observed when  the self-propulsion is increased.  The range of self-propulsions in which this pure liquid regime exists appears to vanish upon approaching the so-called `jamming point' at which solidity of soft-sphere packings is lost.  Such an `active yielding' transition shares similarities with the generic yielding transition for shear flows. A Herschel-Bulkley law is observed along the liquid regime in both loading scenarios, with a clear difference in the critical scaling exponents between the two, suggesting the existent of different universality classes for the yielding transition under different driving conditions. In addition, we present direct measurements of growing length and time scales for both driving scenarios.  A comparison with theoretical predictions from recent literature reveals poor agreement with our numerical results.
\end{abstract}

\maketitle


\emph{Introduction}.---Amorphous materials like dense emulsions, colloids and foams display complex mechanical and rheological responses, which remain mystifying. At large enough density these materials are mechanically stable~\cite{jamming1,jamming2,jamming3,jamming4}, however they can flow uniformly if a sufficiently large stress $\sigma$ is applied~\cite{yield_exp,yield_exp2,yield_exp3,yield_exp_ch}. Growing evidence indicates that this athermal solid-to-liquid transition, known as the~\emph{yielding transition}, displays critical behavior near the yield stress $\sigma_c$~\cite{aval_1}. 

Nowadays it is understood that rheology at low strain rates is controlled by plastic events which are local, irreversible rearrangements of a few tens or hundreds of particles, known as `shear transformation zones'. However, no consensus exists about the essential ingredients controlling the observed constitutive rheological relations. Empirical evidence indicates that rheological flow curves follow the Herschel-Bulkley (HB) law $\dot{\gamma}\!\sim\! (\sigma\!-\! \sigma_c)^\beta$, with  $\dot{\gamma}$ denoting the shear rate, and $\beta$ is the HB exponent (sometimes reported as $n\!=\!1/\beta$), taking values of 2.78~\cite{M_bius_2010} and 3~\cite{yield_exp2} in 2D foams, 1.75~in 3D foams~\cite{beta_exp3},  2.22 for 3D soft colloidal pastes \cite{beta_exp1} and 3D emulsion \cite{beta_exp2}, and 1.81-2.1 for 3D carbopol gel \cite{beta_exp3}. 2D molecular dynamics simulations display a similar HB law with exponent~2.33 \cite{edan_expB}, 2~\cite{beta_sim1},  3.3 \cite{beta_sim2} which seems consistent with experimental measurements.

Several theoretical efforts have been made during the last decade to understand the phenomenon of yielding.  Most of these efforts focus on mesoscopic elastoplastic models \cite{resumen_Eze}, leading to important progresses, including several scaling relation predictions for the critical exponents controlling avalanches and dynamics~\cite{JieLin_1,JieLin_3,elas-HB1,Ezequiel1} and a HB exponent ranging between $\!1\!-\!2.5$ depending on the model details \cite{Ezequiel1,JieLin_3,JieLin_1}. The large dispersion in HB exponents observed from experiment, simulations and theory raise a burning question about the robustness of the critical behavior and/or precision of exponent measurments. 

\begin{figure}[h]
\centering
\includegraphics[width = 0.47\textwidth]{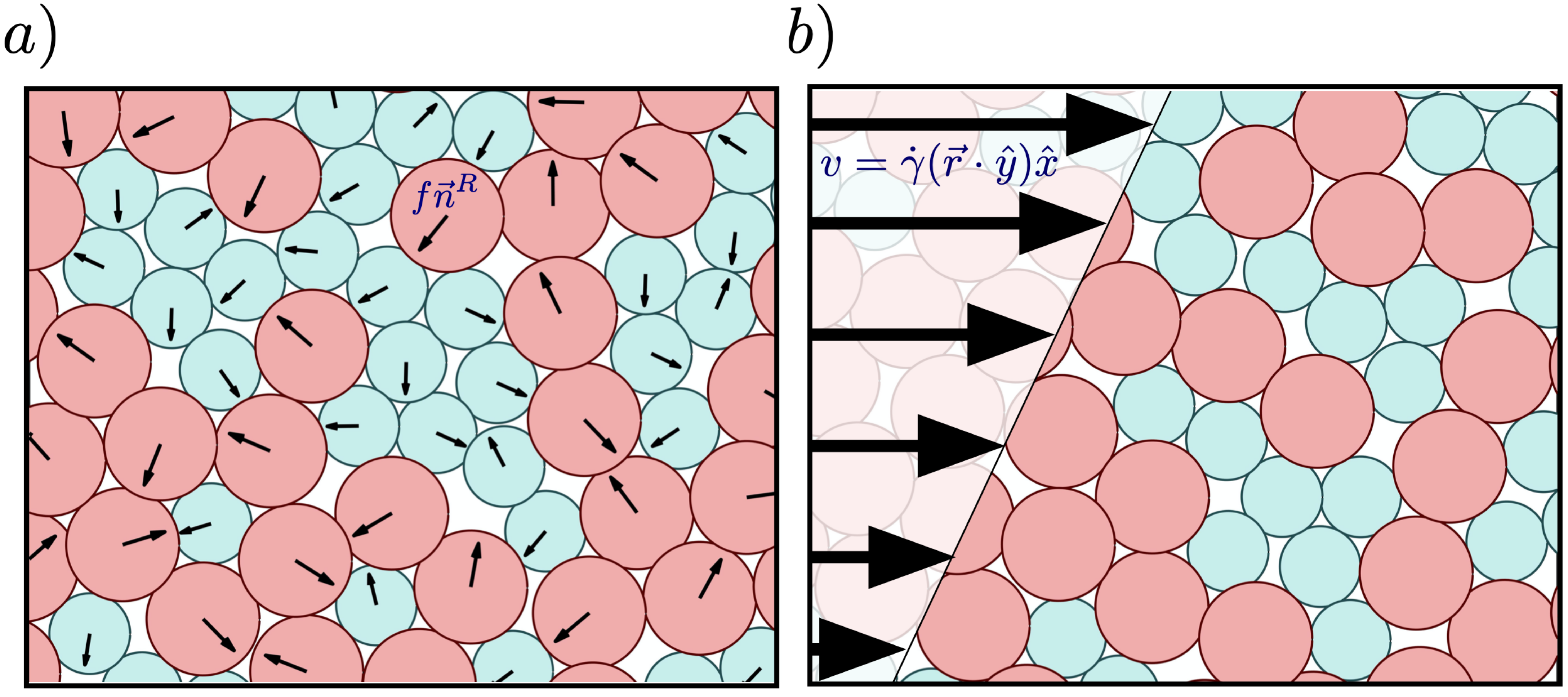}
\caption{\footnotesize a) Self random force model (SRF), the arrow in each particle represents the direction $\vec{n}^R$ of the auto force with size $f$. b) Simple shear  model (SS), where a speed profile $v= \dot\gamma  (\vec{r}\cdot\hat{y})\hat{x}$ is imposed on the system.}
\label{modelos}
\end{figure}
 
Active systems have emerged as a natural system to assess the universality of the yielding transition. In systems of self-propelled particles, such as bacterial colonies or cell tissues, a critical self-propelled force $f_c$ must be overcome in order to initiate persistent flow at sufficiently high densities \cite{ext_ac_in,active_exp2,active_jam}. This transition is expected to be controlled by shear transformation zones similar to sheared amorphous materials, however, results are very scarce~\cite{active_VS_shear}, and the rheology of dense active systems remains largely unexplored. Very recently, inspired in an infinite dimensional model~\cite{inf_dimen}, it has been suggested that flow, deformation and failure in active and sheared amorphous materials can be explained from a universal framework, although only pre-yielding results were tested \cite{active_VS_shear,active_exp1}. 

In this Letter we use extensive numerical simulations of overdamped athermal particle dynamics to carefully study and compare the critical yielding rheology at finite strain rates, under two different driving scenarios; \textbf{Simple shear (SS)} and \textbf{ Self random force (SRF)}, the latter is an active system (see Fig.~\ref{modelos}). We measure the HB exponent for both the passive, SS system, and for the active, SRF system, where in the latter special attention was devoted to avoid the lurking motility-induced phase-separation~\cite{separa_fase2,separa_fase3,separa_fase4}. Remarkably, the HB exponent $\beta$ clearly appears to be different between the two driving scenarios. We confirm these result using two types of interaction potentials and different packing fractions for each scenario;  interestingly, we find no change within each scenario of the HB exponent, suggesting the existence of two distinct universality classes for yielding.  

It has been shown~\cite{cuasi_buble,aval_1,avalan2,avalan3,avalan4} that avalanches grow in size and dynamics becomes more correlated upon approaching the critical yielding point. Therefore, a diverging length scale $\xi\!\sim\! (\sigma\!-\!\sigma_c)^{-\nu}$ that controls the transition is expected to emerge. Unfortunately, only indirect measurement exist for a growing correlation length in SS~\cite{aval_1,edan_expB,largo_in}, and none exist --- to the best of our knowledge --- for SRF. Recently, important finite size effects have been observed that are claimed to introduce difficulties in extracting the critical scaling exponents \cite{yield_exp_ch,buble2}. Therefore, we corroborate our results by direct measurement  of the  correlation length and time scales of avalanche activity. As in the case for the HB exponent $\beta$, $\nu$ also assumes distinct values between SS and SRF scenarios. At the same time, the exponent relating a characteristic correlation time with strain rate remains unchanged between the two scenarios. 

\emph{Models}.---   To avoid crystallization~\cite{cristali_1,cristali_2}, we employ athermal  systems of frictionless soft disks in two dimensions using the standard 1:1.4 bidisperse mixture with equal numbers of disks of two different radii. The radius of the small particles sets the atomistic length scale ($r_0\!=\!1$). Particles interact via a repulsive force with a potential given by
\begin{equation}\label{eq1}
U(r_{ij})=\left \{ \begin{matrix}  \frac{\epsilon}{\alpha} (1-\frac{r_{ij}}{d_{ij}})^\alpha& r_{ij}<d_{ij}
\\ 0 &  r_{ij}>d_{ij}  \end{matrix},\right.\end{equation}
where $r_{ij}$ is the distance between the centers of particles $i$ and $j$, $d_{ij}$ is the sum of their radii, and $\epsilon$ is an energy scale. Two interactions potentials were used, corresponding to $\alpha\!=\!2$ (harmonic) and $\alpha\!=\!5/2$ (Hertzian). For all models, we use overdamped particle dynamics, which follows the equation:
\begin{equation}
\vec{v}_i=\frac{d\vec{r}_i}{dt}=-D\frac{\partial U(r_{ij})}{\partial \vec{r}_i} + \vec{dc_i}\,,
\label{overD1}
\end{equation}
where $\vec{r}_i$ and $\vec{v}_i$ are the position and velocity of particle $i$, respectively, $D$ is the overdamped constant, and time is measured in units of $t_0\!=\! r_0^2/D\epsilon$. Under SS $\vec{dc_i}\! =\!\dot \gamma (\vec{r}_i \cdot \hat{y}) \hat{x}$ which corresponds to setting a velocity profile under direction $\hat{x}$ at a strain rate $\dot \gamma$~\cite{durian_metodo} as shown in Fig.~\ref{modelos}b, where Lees-Edwards boundary conditions~\cite{lees-edwar} are used. For the SRF scenario each particle is subjected to a self-propulsion $\vec{dc}_i\!=\! Df\vec{n}_i^R$, where $f$ is the magnitude of the self-force, and  $\vec{n}_i^R$ is a unit vector with a quenched random direction. Periodic boundary conditions are employed, and a uniform distribution of vector directions is used, while accounting for the constraint $\sum {\vec{n}_i^R}=0$ to avoid center of mass motion. For more details on the numerical methods, see Supplemental Material \cite{SuMa}. 

In the SRF case the natural control parameter is the magnitude of the self force $f$. However, it is more convenient for the study of critical behavior to control the mean parallel velocity $v^R_\parallel\!=\!\frac{1}{N} \sum \vec{v}_i \!\cdot\!\vec{n}^R_i$~\cite{metodos_SRF}. In Supplemental Material \cite{SuMa} we show that both methods lead to the same results, however the constant-parallel-velocity method is more effective in suppressing finite-size effects, and allows for a cleaner comparison with SS. This change is equivalent to controlling the imposed strain rate instead of the imposed shear stress for SS systems~\cite{durian_metodo}.

\begin{figure}[h]
\centering
\includegraphics[width = 0.45\textwidth]{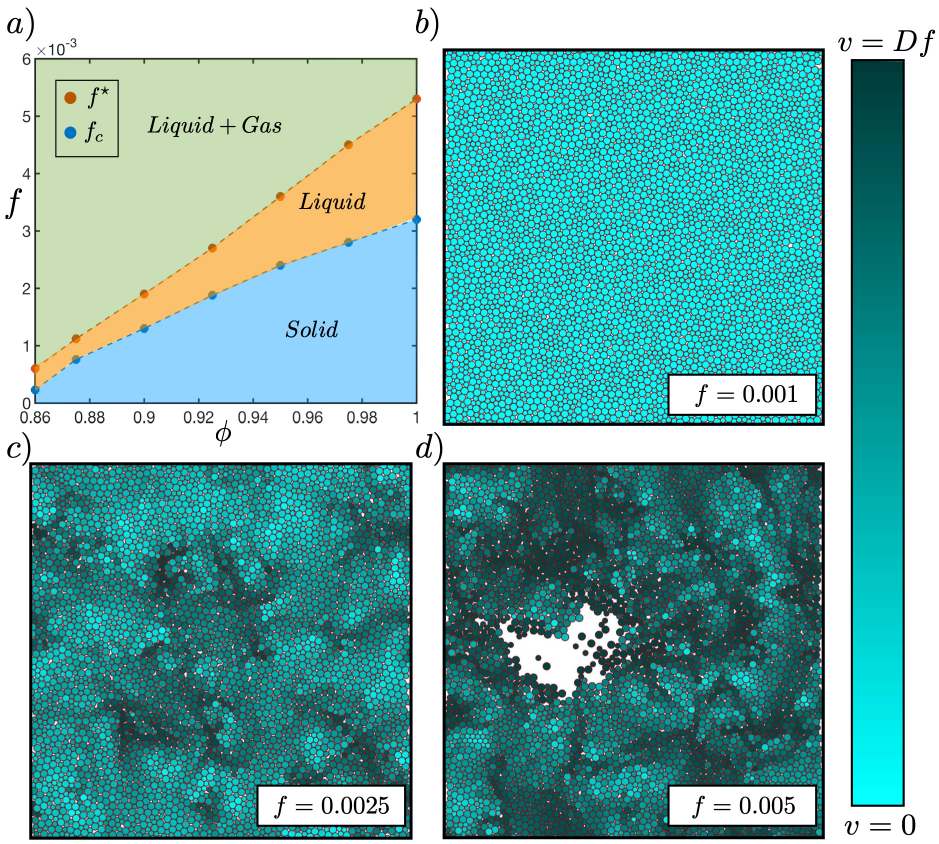}
\caption{\footnotesize a) Phase diagram $f$ - $\phi$  above the jamming point  $\phi_c=0.84$ for the SFR model  with $N\!=\!16384$.  Three different regimes can be identified, which are delimited by $f_c$ and $f^\star$ (see the main text for definitions). b)-d) Activity for different self-force values in a system of $N\!=\!4096$ and $\phi\!=\!0.925$. Darker areas represent particles with larger speed. b) Solid phase, with a self-force value  $f=0.001$.  c) Liquid phase, with a self-force value  $f=0.0025 $.  It is not possible to find a state of equilibrium $ f> f_c $, the movement is associated with correlated avalanches. d) Phase separation, with a self-force value $f=0.005$ it is possible see nucleation and a gas phase area with free particles. The color scale goes from $|\vec{v}_i|=0$ to $|\vec{v}_i| = Df$ for the movement of a free particle. }
\label{fases}
\end{figure}


\emph{SRF phase diagram}.--- To our knowldege, the phase diagram for SRF has not been extensively explored at large densities in previous work. As expected, above Jamming and for small enough self-propulsion force the material remains in a solid phase (see Fig.~\ref{fases}b). Above a critical magnitude $f_c(\phi,N)$ the system yields and no mechanical equilibrium configurations are found. The critical force $f_c$ is analogous to the yield stress $\sigma_c$ in the SS Herschel-Bulkley dynamics. A finite-size analysis shows that, in the thermodynamic limit, $f_c(\phi,N\rightarrow \infty)$ seems to converge to a finite value (see Supplemental Material \cite{SuMa}), validating the existence of an active yielding transition. Above $f_c$ the system is in a continuous process of reorganization dominated by avalanches triggered at STZs. Fig.~\ref{fases}c shows high-speed areas associated with existence of heterogenous avalanches~\cite{avalan2}.

\begin{figure}[h]
\centering
\includegraphics[width = 0.45\textwidth]{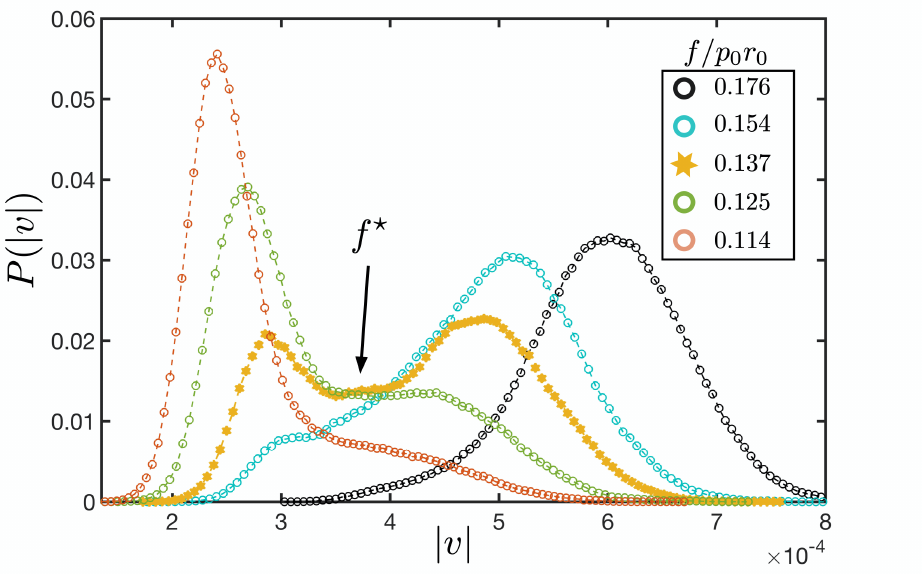}
\caption{\footnotesize Mean velocity distribution $P(|v|)$ for $N=16384$, $\phi=0.925$ and harmonic potential. The data corresponding to yellow hexagrams show a bimodal distribution, where the first peak at low mean velocity corresponds to purely liquid configurations, and the second peak at high mean velocity corresponds to a configuration with nucleation; this bimodal distribution defines $f^\star$ in this method ($ f^\star/p_0r_0 = 0.137 $). For $f>f^\star$ there is only one peak, indicating that most of the configurations exhibit nucleation. Similarly, $f<f^\star$ there is only one peak, indicating that most of the configurations are in liquid regime.}
\label{distri}
\end{figure}

Motility-induced phase separation in active matter is a recurrent phenomenon~\cite{separa_fase2,separa_fase4}. For large enough $f$ the SRF system displays void-nucleation and gas-liquid phase-coexistence, as observed in Fig.~\ref{fases}d. The ocurrence of this phenomenon hinders the observation of the HB law, since its validity can only be assured in homogeneously flowing liquid phases. The transition from the homogeneous liquid phase and a regime with bubble nucleation can be set at $f^\star$ by looking at the mean velocity  of the system $|v|= \frac{1}{N} \sum_i^N \sqrt{\vec{v}_i\cdot\vec{v}_i}$ at different times.   In Fig.~\ref{distri}, we show the  distribution $P(|v|)$ of mean velocities at different  $f$. From the distribution we can see the bubble nucleation from the appearance of a second peak at much larger velocity, which is controlled by  particles in the gas phase. We set $f^\star$ as the self-propulsion for which the homogeneous liquid state and the regime with nucleation are equally likely. An alternative, and more easy way to establish the transition is by the pressure, which shows an abrupt increase at $f^\star$ (see Supplemental Material \cite{SuMa}). Both methods lead to  similar results.


In Fig.~\ref{fases}a, we present the obtained phase diagram with the three aforementioned regions.  The existence of pure gas phase could take place at higher forces, but was not considered in this work. We note that the uniform liquid phase region appears to vanish when approaching the Jamming transition at $\phi_J=0.843$ \cite{jamming4,jamming5}, suggesting that rheology at the Jamming packing fraction could be of different nature.


\begin{figure}[h]
\centering
\includegraphics[width = 0.47\textwidth]{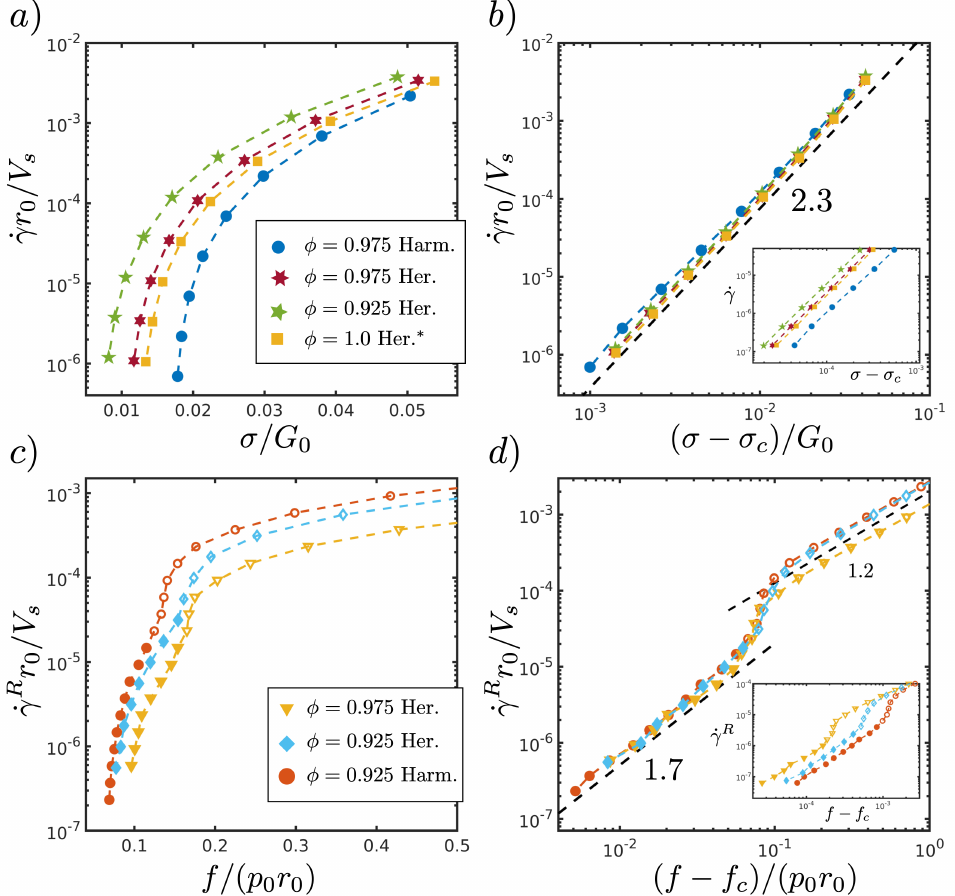}
\caption{\footnotesize  Flow curves  a) Shear rate $\dot{\gamma}$ vs. shear stress  $\sigma$ for SS.  Four different configurations of $\phi$ and $N=65536$ (the data with ${}^*$ $N=16384$ ) were used. b) Same data, shear rate and the difference between shear stress and yield stress on log scales (inset: raw data).  c) Equivalent shear rate $\dot{\gamma}^R$ vs. self-force $f$ for SRF model, three different configurations and $N=16384$ were used. d)  Same data in equivalent shear rate $\dot{\gamma}^R$ vs.~the difference between self-force and critical self-force $f_c$ (inset: raw data).}
\label{HB_1}
\end{figure}

\emph{Flow curves}.--- To compare the HB exponents in the SS and SRF scenarios, we define the strain rate $\dot{{\gamma}^R}$ for SRF model using the parallel velocity $v_\parallel^R$ defined above. In the case of SS one can easily shown from Eq.(\ref{overD1}) that $v_\parallel^S \! = \!\frac{D}{N}  \sum_i\frac{\partial U(r_{ij})}{ \partial \vec{r}_i} \cdot \hat{x} \!+\!  \frac{\dot \gamma}{N}  \sum_i \vec{r}_i \cdot \hat{y}$, where the sum of contact forces is zero due to boundary conditions. Therefore, we obtain $v_\parallel^S \! = \!  \dot{\gamma} L/(2\sqrt{N})$, which leads to the definition $\dot{{\gamma}^R}\!=\! \frac{2\sqrt{N}}{L} v_\parallel^R$~\cite{active_VS_shear}.

\begin{figure}[h]
\centering
\includegraphics[width = 0.48\textwidth]{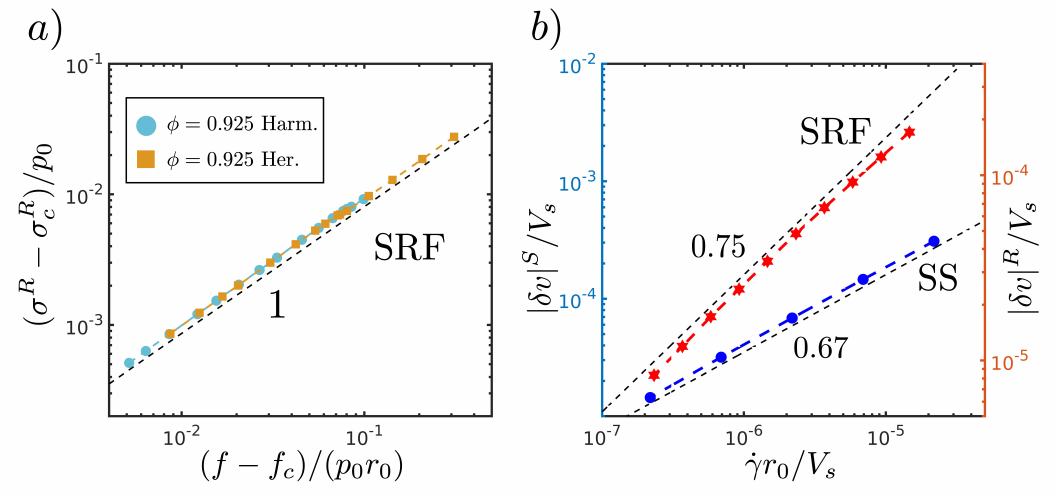}
\caption{\footnotesize a) Relation between  ``random" stress and self force i.e. $\sigma^R- \sigma^R_c$ vs.  $f-f_c$, for $N\!=\!16368$, $\phi\!=\!0.925$, and both potentials. b) Mean square root velocity fluctuations vs. shear rate $\dot{\gamma}$ for both models, with $N=16348$, $\phi=0.925$ and using the Hertzian potential. }
\label{HB_2}
\end{figure}

In Fig.~\ref{HB_1}a,b we present the SS model flow curves, where we find a HB law $\dot{\gamma}\!\sim\! (\sigma\!-\!\sigma_c)^\beta$ with $\beta\!\approx\!2.3$, the latter remains unchanged for different packing fractions and interaction potentials. This measured exponent is in agreement with some previous results with similar simulation protocols~\cite{edan_expB}. We further note that curves collapse  when rescaled as $\sigma\!\rightarrow\!\sigma/G_0$ and $\dot{\gamma}\!\rightarrow\!\dot{\gamma} r_0/V_s$, where $G_0$ is the shear modulus in the pre-yielded system, and $V_s\!=\!\sqrt{G_0/\rho}$ is the speed of sound, with $\rho$ denoting the mass density. For more details of simulations protocols and fitting parameters, see Supplemental Material \cite{SuMa}.

The SRF scenario requires more care. In Fig.~\ref{HB_1}c,d we identify two different regimes; the first one corresponds to the phase separation region. A large contribution to $\dot{\gamma}^R$ of the particles in a gaseous state explains the abrupt jump in the strain rate. The real yielding transition takes place in the pure liquid regime --- represented by the filled points in Fig.~\ref{HB_1} --- leading to a HB law $\dot{{\gamma}^R}\!\sim \!(f\!-\! f_c)^{\beta}$ with $\beta\!\approx\!1.7$, independent of packing fraction and interaction potential. This result shows a remarkable difference in the $\beta$-exponent between SS and SRF models. Again, our data collapse very well considering the dimensionless quantities $\dot{\gamma}^R r_0/V_s$ and $f/(p_0 r_0)$. 

We note that a stress field for the SRF model can be defined using a deformation about the direction of the random vector, given by~$\vec{r}_i \rightarrow \vec{r}_i  + \frac{L}{2\sqrt{N}} \gamma^R \hat{n}^R_i$. Then the ``random'' stress field that is equivalent to the shear stress for the SS scenario is given by $$\sigma^R=\frac{1}{L^2} \frac{dU}{d\gamma^R}  =\frac{1}{2 L \sqrt{N}} \sum_{i=1}^N  \frac{\partial U}{\partial \vec{r}_i}   \hat{n}^R_i.$$
Combining the last result and Eq.~(\ref{overD1}) one obtains the relation $\sigma^R\!=\! \frac{\sqrt{N}}{2L}f\!-\!\frac{1}{4D}\dot{\gamma}^R$.  Since $\dot{\gamma}$ decays to zero faster that $f\!-\! f_c$ when approaching the transition, one obtains that $\sigma^R\!-\!\sigma^R_c\!\sim\! f\!-\! f_c$ with $\sigma^R_c\!\equiv\!\frac{\sqrt{N}}{2L}f_c$ as observed in Fig.~\ref{HB_2}a.

\begin{figure*}[t]
\centering
\includegraphics[width = 1\textwidth]{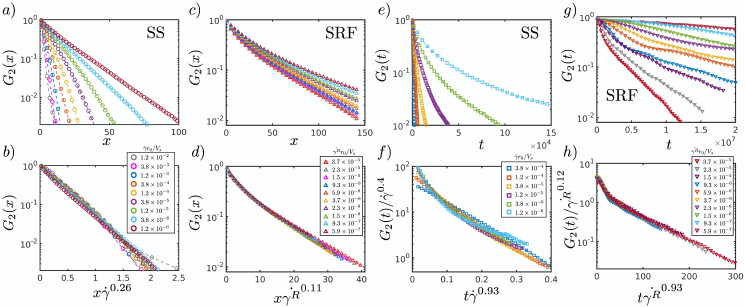}
\caption{\footnotesize a) Spatial correlation function $G_2(x)$ for SS at different strain rates. b) Same data of a) after a proper rescaling. c) Spatial correlation function for SRF at different strain rates. d) Same data of c) after a proper rescaling.  e) Time correlation function $G_2(t)$ for SS at different strain rates. e)  Same data of a) after a proper rescaling.  g) Time correlation function $G_2(t)$ for SRF at different strain rates. h) Same data of g) after a proper rescaling. The data correspond to Hertianz potential, $\phi=0.925$ and $N=65536$ for SS and $N=16348$ for SRF. All correlation are normalized such that $G_2(x=0)=1$ and $G_2(t=0)=1$.}
\label{CE_1}
\end{figure*}

An alternative route towards quantifying the activity of avalanches is provided by studying fluctuations around the mean parallel velocities. For SS this corresponds to the standard `non-affine' velocity $\delta \vec{v}_i\!\equiv\!\vec{v}_i\!-\!\dot{\gamma}(\vec{r}_i\!\cdot\!\hat{y})\hat{x}$, and for SRF we define the fluctuations as $\delta \vec{v}_i\!\equiv\!\vec{v}_i\!-\!v_\parallel^R \hat{n}_i^R$. The mean square root velocity fluctuations are shown in Fig.~\ref{HB_2}b; an important difference is observed in the exponent where $|\delta v^S|\!\sim\!\dot{\gamma}^{0.67}$ for SS, and $|\delta v^R|\!\sim\!\dot{\gamma^R}^{0.75}$ for SRF. Interestingly, the avalanche activity seem to increase much faster with strain rate for SRF.

\emph{Correlations}.--- A key phenomenon in the yielding transition is that dynamics becomes more cooperative and dominated by avalanches that grow in size close to the yielding point \cite{aval_1,avalan3}. This cooperative dynamics should be controlled by a diverging lengthscale which so far has not been measured directly in molecular dynamics simulations~\cite{correla_2,aval_1}. A correct measurement of a growing correlation length is essential to validating the HB $\beta$ exponent, since several strong finite size effects have been suggested to exist close to yielding~\cite{cuasi_buble,avalan2}. 

Olsson \cite{asy_O} reported that the computational calculation to obtain the correlation length and resolve its scaling behavior is an arduous numerical task; he further comments that the two-point correlation associated with non-affine velocity for a system under shear presents zones of correlation and anticorrelation associated with the angular orientation in which the correlation is measured. Therefore to avoid anticorrelation effects in the orientation dynamics, we propose to measure the correlation using the spatial activity field of the non-affine velocity, defined as
\begin{equation}
G_2(x)=\langle  |\delta \vec{v}(0)|| \delta \vec{v}(x)| \rangle-\langle  |\delta \vec{v}(0)| \rangle\langle | \delta \vec{v}(x)| \rangle.
\end{equation}
This correlation function is inspired from the idea that avalanches generate high velocity fluctuation zones whose size can be estimated by neglecting the direction in which the movement is executed. Our proposition is very similar to the correlation put forward by Hurley and Harrowell to measure sizes of structures in equilibrium liquids \cite{idea_Corre}.

In Fig.~\ref{CE_1}a-d we present the correlation function for both SS and SRF scenarios. A clear exponential decay is observed, allowing us to directly extract the correlation length $\xi\! \sim\! \dot{\gamma}^{-0.26}\!\sim\! (\sigma\! -\! \sigma_c)^{-0.6}$ for the SS model, and $\xi^R\!\sim\!\dot{\gamma^R}^{-0.11}\!\sim\!(f\! -\! f_c)^{-0.19}$ for the SRF model.  A clear collapse is observed for both models.  We calculated the correlation length for different packing fraction and interaction potential, and the exponents remain unchanged for each type of model.

The high velocity fluctuation zones have a lifetime that sets a correlation timescale. One can associate with this timescale the time  elapsed between a plastic event is triggered and until its effect is no longer seen in the system. In a completely cooperative dynamics, this time scale should diverge in the thermodynamic limit. To measure the exponent associated with this timescale we define the autocorrelation
\begin{equation}
G_2(t)=\left \langle  \frac{\delta \vec{v}(0) \cdot \delta \vec{v}(t)}{|\delta \vec{v}(0)||\delta \vec{v}(t)|}   \right \rangle.
\end{equation} 
Here we only use the direction of the non-affine velocity. This expression allows us to avoid anticorrelations which seem to be caused by a strong suppression of velocity fluctuation after an avalanche's end, and to define the average lifetime of an avalanche in the system $t^* \sim \dot{\gamma}^{-\eta}$. As shown in Fig.~\ref{CE_1}e-h, unlike the previous exponents reported above, $\eta$ remains unchanged between the SS and SRF models, and found to be $\eta\!\approx\!0.93$. We have corroborated this exponent for SS using an oscillatory shear method described in Supplemental Material \cite{SuMa}. Notice importantly that in order to measure a correct HB exponent we consider a time average over a time scale much longer than this correlation time scale.

\emph{Summary}.---In this Letter we have shown that the exponents of the HB law and the correlation length scale are different between the SS and SRF scenarios, however remain invariant to changing packing fraction or interaction potential. Neither the mean field model proposed by Haubraud and Lequeux \cite{mean_field}, nor more elaborated elastoplastic models \cite{elasto1,JieLin_2,JieLin_3,elasto2,Ezequiel1}, seem to capture the observed dependence on the nature of the driving geometry, and is thus new physics. The origin of our observed difference between the two scenarios is unknown to us, but should be key to build more accurate models and predictions.

Inspired by the depinning transition, a set of scaling relation were obtained for the yielding transition~\cite{JieLin_2}. The correlation lengthscale exponent is suggested to be $\nu\!=\!1/(d\!-\! d_f)$ where $d$ is the spatial dimension and $d_f$ is the fractal dimension extracted from the avalanche distribution in the quasi-static regime at yielding. Considering the exponent $\nu$ extracted from our data, the fractal dimension should be $d_f\!\approx\!0.3$ for simple shear, which is very different from previous values observed in molecular dynamics that were close to 1~\cite{df_1,df_2}. 

 Another key exponent in order to understand the scaling relation~\cite{exp_z,JieLin_3} is the one relating the linear size of an avalanche $l$ with the duration $T$ in which such avalanche takes place, such that $T\!\sim\! l^z$. The scaling relation put forward in~\cite{JieLin_2} gives $z\!=\!(\beta\!-\!1)/\nu$; using our measured $\beta$ and $\nu$ exponents we obtain $z\!\approx\!2.2$, which is much larger that the one observed in elastoplastic models~\cite{JieLin_1,Ezequiel1}. In addition, for our system at finite strain rate one could expect that the correlation timescale and the correlation lengthscale follow a scaling relation $t^*\sim \xi^z$ with the same exponent $z$. However, we observe that $t^*\sim \xi^{\!\beta \eta/\nu\!}\sim  \xi^{3.5}$ for SS, which is much larger exponent than the one predicted by the scaling relation. An immediate future research direction is a revision of the scaling relations for molecular dynamics simulations, and more importantly, a detailed study of the quasi-static dynamics at yielding for SS and SRF, in order to ascertain if the avalanche statistics also features important differences stemming from the nature of the driving.

\textbf{ Acknowledgment.} We thank  Edan Lerner, for fruitful discussions and for comments on the manuscript.  G.D. acknowledge funding from Millennium Science Initiative of the Ministry of Economy, Development and Tourism, grant ``Nuclei for Smart Soft Mechanical Metamaterials". C.V. acknowledges support from ANID. for the scholarship No. 21181971, BN.

\bibliography{Biblio1}
\appendix
\newpage

\section{Simulation protocols.}
\label{Appen1}

\begin{figure}[h]
\centering
\includegraphics[width = 0.35\textwidth]{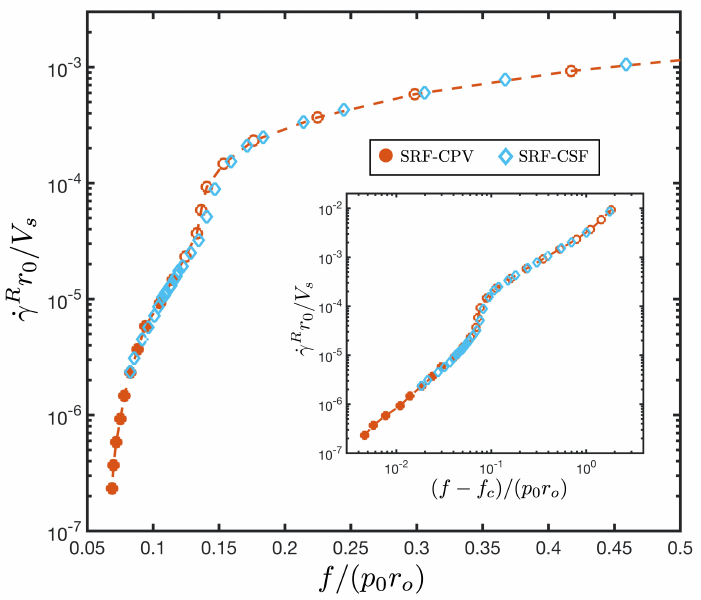}
\caption{\footnotesize $\dot{\gamma}^R$ vs. $f$ using SRF-CPV and SRF-CSF (section without effects of finite size) for $N=16384$, $\phi=0.925$ and harmonic potential. It is observed that both curves show the same behavior.}
\label{Anexo1_1}
\end{figure}

Here we give more details about the simulation protocols and how the data analysis is done for both models. In SS models, like the one proposed by Durian \cite{durian_metodo}, a speed profile is generally imposed with a given shear rate $\dot{\gamma}$ value, and as a consequence, the value of $\sigma$ is measured as the temporal average over a sufficiently long simulation time. 
In SRF model, we distinguish two different kind of algorithm \cite{metodos_SRF}, the first at constant self-force (SRF-CSF), where the value of $f$ remains fixed throughout the simulation, and the second at constant parallel velocity (SRF-CPV), where the value $v^R_\parallel=\frac{1}{N} \sum \vec{v}_i \cdot \vec{n}^R_i$ is set, and in each step the value of $f$ is adjusted to keep $v^R_\parallel$ constant. The SRF-CPV method has the advantage of allowing exploration closer to the critical point. This is because, similarly to what happpens in systems under shear, it is possible for the system to find equilibrium for $f>f_c$ due to finite size effects \cite{buble2,yield_exp_ch}. In this line, the SRF-CPV method ensures a flow that does not suddenly stagnate in the vicinity of $f_c$. In Fig.~\ref{Anexo1_1} we show that the choice of simulation method does not change the results in areas where sudden stagnation is not observed.

One last point to discuss is how the value of $\sigma_c$ and $f_c$ are calculated.  For SS and SRF-CPV, the values are calculated by fitting the HB curve and looking for the values of $\sigma_c$ and $f_c$ that maximize $\chi^2$, respectively.  In the case of SRF-CSF, a sweep is made in $f$ values, with a jump of $\Delta f = 0.00005 $, and $ f_c $ is set for the highest value at which stagnation is appreciated. We also verify that calculation of $ f_c $ is independent of the simulation method.
\\
 To simulate the system dynamics, the time is measured in units of $t_0=r_0^2/D\epsilon$, and we integrate the overdamped particle equation using the RK-2 method at each time step. Unless otherwise specified, we use two system sizes $N=16384$ and $N=65536$ with  $\Delta t=0.85t_0$, where we have verified that the selection of $\Delta t$ does not affect results. For all our results using the SS model, average is taken over 20 different configurations, and for all our results using the SRF model, average is taken over 96 different configurations.
 
 
 \section{Irving-Kirkwood}
\label{Appen2}

To quantify pressure and shear stress we use the Irving-Kirkwood calculation \cite{stress_form} for the stress tensor $\sigma_{\alpha \beta}$. We do this to avoid neglecting the effects of free particles that may be present in gaseous areas.

\begin{equation}
\sigma_{\alpha \beta}=\frac{1}{V}\left( \sum_{i<j} \vec{r}_{ij,\alpha}  \vec{f}_{ij,\beta} + \sum_i  \delta\vec{v}_{i,\alpha} \delta \vec{v}_{i,\beta}  \right). 
\end{equation}

In this equation, the indices $\alpha$ and $\beta$ are the cartesian coordinates, $\vec{f}_{ij}$ and $\vec{r}_{ij}$ are vector force and vector distance between the particles, $\delta \vec{v}_i$ are  the fluctuations around the mean parallel velocities, which we define as $\delta \vec{v}_i=\vec{v}_i-\dot \gamma (\vec{r}_i \cdot \hat{y}) \hat{x}$ for SS and $\delta \vec{v}_i=\vec{v}_i-v_\parallel \hat{n}_i^R$ for SRF. With these expressions, we seek to cancel out the contributions of the deformations to the velocity $\vec{v}_i$. In order to do this in the SS model, we need only to subtract the speed profile term. On the other hand, for the SRF model, we know that the effect of the self-force will lead to each particle moving with an mean velocity $v_\parallel$; for this reason, we consider that the vectorial term that provides the deformation can be written as $v_\parallel \hat{n}_i^R$. Shear stress $\sigma$ is defined as $\sigma\equiv\sigma_{xy}$, and the pressure as $p\equiv (\sigma_{xx}+\sigma_{yy})/2$.

\newpage
\section{$f_c$ at infinity system size.}
\label{Appen3}
\begin{figure}[h]
\centering
\includegraphics[width = 0.48\textwidth]{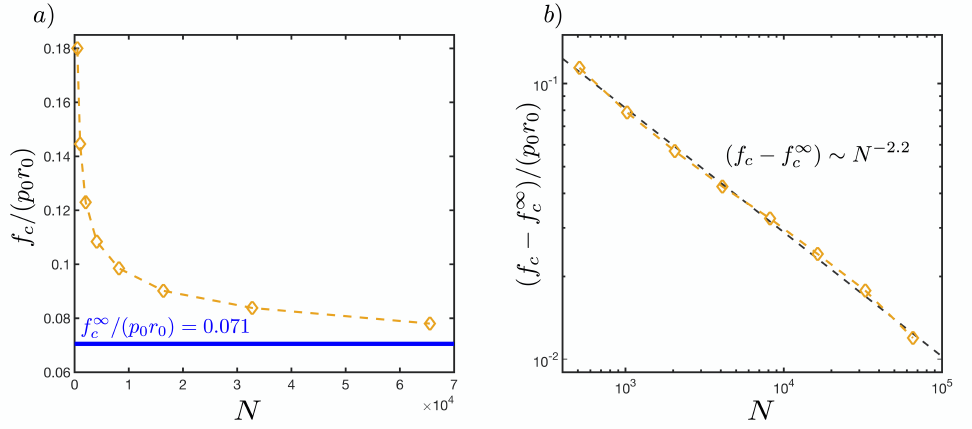}
\caption{\footnotesize For $\phi=0.925$ and harmonic potencial: a) $f_c$ vs. $N$, the data appears to converge to $f_c^\infty/p_0r_0 = 0.071$. b) $f_c-f_c^\infty$ vs. $N$, the data shows a power law $(f_c(N)-f_c^\infty)\sim N^{-2.2}$.}
\label{Anexo2_1}
\end{figure}

Here, we do a finite-size analysis, which is essential to validate the existence of an active yielding transition in the thermodynamic limit $f_c(\phi,N\rightarrow \infty)$. Using the SRF-CSF, we calculate how the value of $f_c$ depends on $N$. These results are shown in image Fig.\ref{Anexo2_1}a, where it can be seen that the value of $f_c$ saturates at a value $f_c^\infty$ in the limit $N\rightarrow  \infty$.

For our simulation with $ \phi = 0.925 $, $ f _c^\infty/p_0r_0=0.071$ is obtained, and a power law  $ (f_c(N)-f_c^\infty)\sim N^{-2.2}$ is appreciated (see Fig.~\ref{Anexo2_1}b). The existence of a non-zero value of $f_c^\infty$ suggests that the presence of an active yielding transition is not a finite size problem.
\newpage
\section{Phase separation detection.}
\label{Appen4}

\begin{figure}[h]
\centering
\includegraphics[width = 0.48\textwidth]{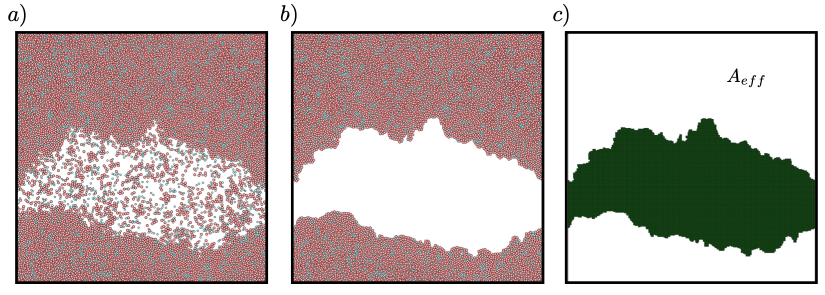}
\caption{\footnotesize Tessellation algorithm scheme for $N=16384$, $\phi=0.925$: a) system that exhibits a nucleation zone; b) system where all particles with less than three contacts were removed, effectively removing the gaseous zone from the system; c) the area where the removed particles were (in green) is computed using the tessellation algorithm, and this is used to calculate the effective area $A_{eff}$.}
\label{Anexo4_2}
\end{figure}

\begin{figure}[h]
\centering
\includegraphics[width = 0.38\textwidth]{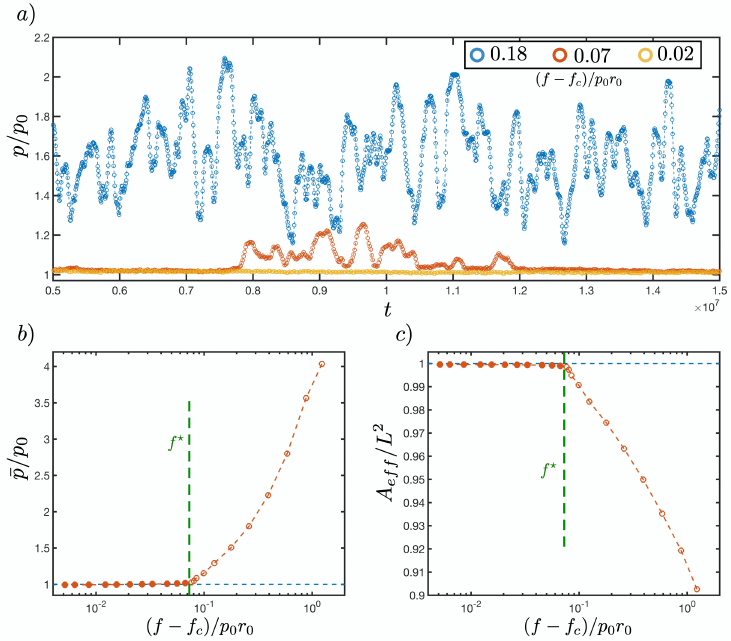}
\caption{\footnotesize For $N=16384$, $\phi=0.925$ and harmonic potential: a) Pressure $p$ vs. simulation time $t$. For $(f-f_c)/p_0r_0=0.02$, no increase in pressure is observed (homogeneous liquid). For $(f-f_c)/p_0r_0=0.07$ an increase in pressure is observed in some instances (there are some configurations where nucleation occurs). For $(f-f_c)/p_0r_0=0.18$, an increase in pressure is observed for all simulation times (all configurations show nucleation). b) Mean pressure $\bar{p}$ vs. $(f-f_c)/p_0r_0$. c) Effective area $A_{eff}$ vs. $(f-f_c)/p_0r_0$. }
\label{Anexo4_3}
\end{figure}

A second detection method, which requires less computational effort, is based on the idea that when nucleation is present, the effective area $A_{eff}$ occupied by the particles in liquid areas decreases, which causes an increase in the global pressure $p$. For the calculation of the effective area $A_{eff}$, a tessellation algorithm on free space is used (see Fig.~\ref{Anexo4_2}), adjusted to have a maximum error of $1.5\%$. The Fig.~\ref{Anexo4_3}a shows how $p$ evolve over the simulation time. 

In Fig.~\ref{Anexo4_3}b,c we see how, for values of $f>f^\star$, there are sudden increases in pressure $p$ which overlap with a decrease in $A_{eff}$.  In our data, we consider that a system does not show phase separation when, for a given value of $f $, it  satisfies the condition $p/p_0 <1.05 $ for all configurations at all simulation times.

\newpage
\section{Shear oscillation.}
\label{Appen5}

\begin{figure}[h]
\centering
\includegraphics[width = 0.44\textwidth]{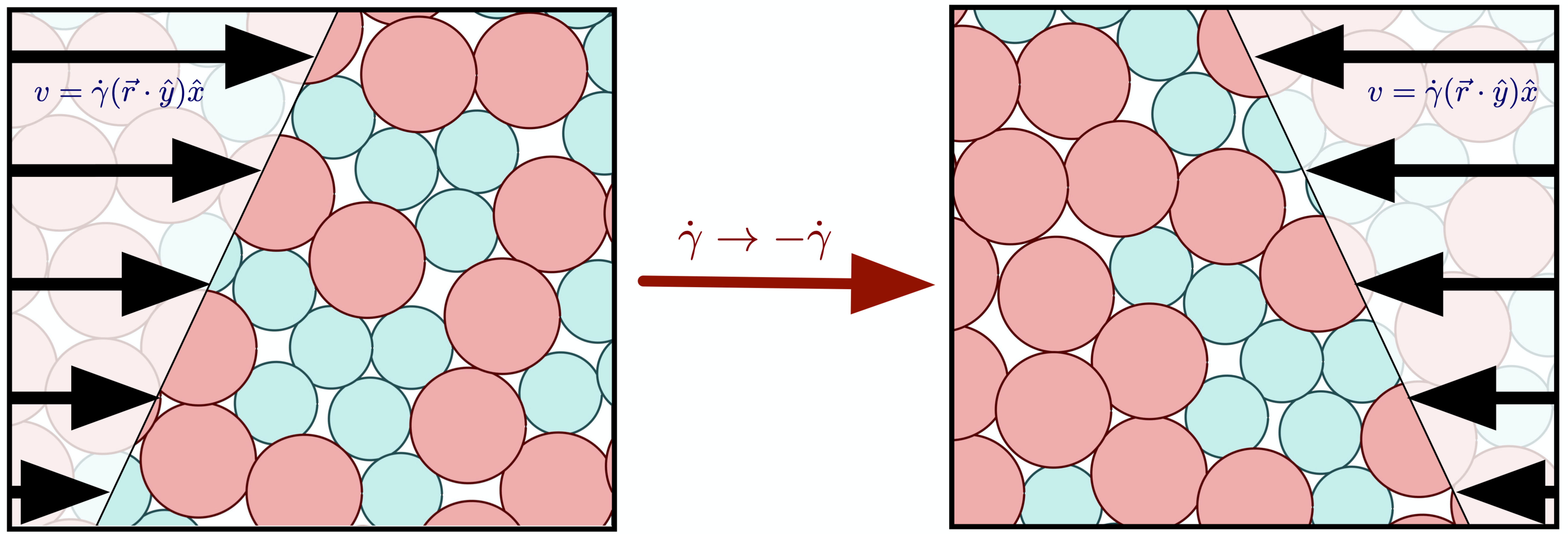}
\caption{\footnotesize Shear oscillation scheme. The orientation of the velocity profile changes abruptly ($\dot{\gamma}  \rightarrow -\dot{\gamma}$). }
\label{Anexo5_1}
\end{figure}

The existance of a correlation time related to the structure's lifetime can also be verified using oscillation simulations for the SS model. Here we invert the orientation of the velocity profile $\dot{\gamma} \rightarrow -\dot{\gamma}$, which causes a change in the measured stress value $\sigma \rightarrow -\sigma$ (see Fig.~\ref{Anexo5_1}). Due to the presence of  these structures, the jump between $\sigma$ and $-\sigma$ is not instantaneous, so we define the lifetime of the structures $t^*$ as the time necessary for this change in the stress value to occur.

\begin{figure}[h]
\centering
\includegraphics[width = 0.46\textwidth]{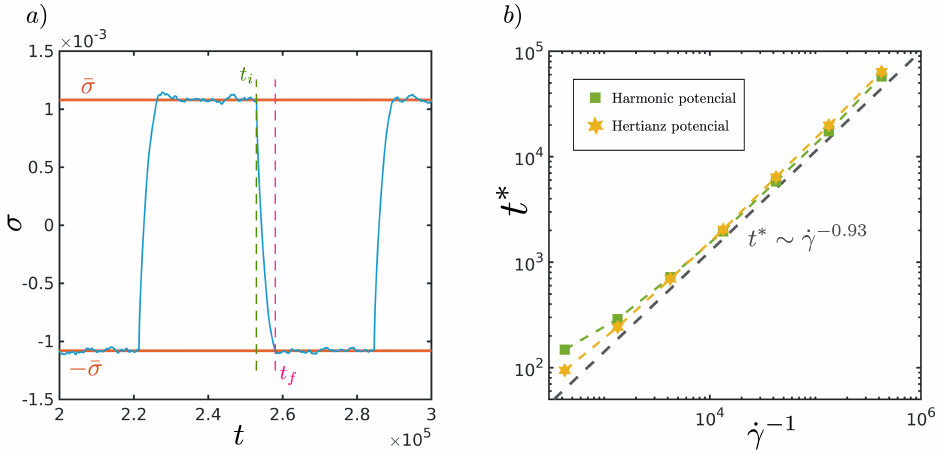}
\caption{\footnotesize For shear oscillation method: a)  Shear stress $\sigma$ vs. simulation time $t$ for $N=65536$ and  $\dot{\gamma}=3.2\times 10^{-5}$; shear stress changes its orientation from $\bar{\sigma}$ to $-\bar{\sigma}$ between times $t_i$ and $t_f$. b) $t^*$ vs. $\dot{\gamma}^{-1}$ for $N=65536$ and both potentials, a power law $t^*\sim \dot{\gamma}^{-0.93}$ is obtained. }
\label{Anexo5_2}
\end{figure}

The Fig.~\ref{Anexo5_2}a shows how this process is carried out. Firstly, the system is subjected to a shear rate $\dot{\gamma}$, and as a result, the stress varies around an average value $\bar{\sigma}$. Then, in time $t_i$, the orientation changes abruptly to $-\dot{\gamma}$, and we wait until the value $-\bar{\sigma}$ is reached in time $t_f$. In Fig.~\ref{Anexo5_2}b  we show how $t^*=t_f-t_i$ depends on  $\dot{\gamma}$ ,which verifies that the lifetime of the structures in the SS simulations follows a power law $t^*\sim \dot{\gamma}^{-0.93}$.
\\


\end{document}